\documentclass[aps,preprint,prl,twocolumn,tightenlines,10pt,showpacs]{revtex4}%

\usepackage{amsfonts}
\usepackage{amsmath}
\usepackage{amssymb}
\usepackage{graphicx}%
\begin{document}
\preprint{M.\ Durand}
\title[Minimum cost pipe network/Optimal transport and irrigating networks,
irrigation networks. Efficient constrained transport networks]{Structure of optimal pipe networks subject to a global constraint}
\author{Marc Durand}
\affiliation{Mati\`{e}re et Syst\`{e}mes Complexes, UMR 7057 CNRS \& Universit\'{e} Paris 7
- Denis Diderot, Tour 33/34 - 2\`{e}me \'{e}tage - case 7056, 2 Place Jussieu
- 75251 Paris Cedex 05, France }

\begin{abstract}
The structure of pipe networks minimizing the total energy dissipation rate is
studied analytically. Among all the possible pipe networks that can be built
with a given total pipe volume (or pipe lateral surface area), the network
which minimizes the dissipation rate is shown to be loopless. Furthermore,
such an optimal network is shown to contain at most $N-2$ nodes in addition to
the $N$ sources plus sinks that it connects. These results are valid whether
the possible locations for the additional nodes are chosen freely or from a
set of nodes (such as points of a grid). Applications of these results to
various physical situations and to the efficient computation of optimal pipe
networks are also discussed.

\end{abstract}
\volumeyear{2006}
\volumenumber{number}
\issuenumber{number}
\eid{identifier}
\date[Date text]{date}
\received[Received text]{date}

\revised[Revised text]{date}

\accepted[Accepted text]{date}

\published[Published text]{date}

\startpage{1}
\endpage{2}
\maketitle

Finding the most efficient transport network is an issue arising in a wide
variety of contexts \cite{Stevens,Ball}. One can cite, among others, the
water, natural gas and power supply of a city, telecommunication networks,
rail and road traffic, and more recently the design of labs-on-chips or
microfluidic devices. Moreover, this problem also appears in theoretical works
intending to describe the architecture of the vascular systems of living
organisms \cite{McMahon,Durand}. Generally speaking, consider a set of sources
and sinks embedded in a two- or three-dimensional space, their respective
number and locations being fixed. The flow rates into the network from each
source, and out of the network through each sink, are also given. The problem
consists in interconnecting the sources and sinks via possible intermediate
junctions, referred to as \textit{additional nodes}, in the most efficient
way. That is, to minimize a cost function of general form $\sum_{k}%
w_{k}f\left(  i_{k}\right)  $, where the summation is over all the links that
constitute the network. $w_{k}$ is the \textquotedblleft
weight\textquotedblright\ associated with the $k$th link, and $f$ is some
function of the flow rate $i_{k}$ carried by this link. Minimization of the
cost function can be done over different optimization parameters and with
different constraints.

Here, the structure of pipe networks that minimize the total
\textit{dissipation rate} $U=\sum_{k}r_{k}i_{k}^{2}$ is studied, where the
weight $r_{k}$ is the \textquotedblleft flow resistance\textquotedblright\ of
pipe $k$, defined as:%
\begin{equation}
r_{k}=\frac{\rho l_{k}}{s_{k}^{m}}, \label{resistance}%
\end{equation}
$\rho$ being some positive constant, $l_{k}$ and $s_{k}$ the length and
cross-sectional area of each pipe respectively, and $m$ a positive constant
characterizing the flow profile. For most flows encountered in physics,
$m\geq1$ (some examples of flows are given later). In this letter, two major
results are reported. First, among all the possible pipe networks that can be
built with a given value of total pipe volume (or total lateral surface area),
the network that minimizes $U$ is loopless. This result suggests an
explanation for the observed topologies in the vascular systems of various
living organisms \cite{McMahon,Banavar}. Second, the number of additional
nodes in this optimal network cannot exceed $N-2$, where $N$ is the number of
initial nodes (sources plus sinks). As a consequence, the number of possible
different topologies for the optimal network is finite.

Flow rates in a network are not independent but must satisfy a conservation
law at every source, sink, and additional node. That is, the sum of algebraic
flow rates at each site must satisfy:
\begin{equation}
\sum_{\substack{adjoining\\pipes}}i_{k}=%
\genfrac{\{}{.}{0pt}{}{0\text{ at every additional node}}{I_{q}\text{ at
source or sink }q}
\label{kirchhoff}%
\end{equation}
where $I_{q}$ is the fixed inflow/outflow at the source/sink $q$ ($I_{q}>0$
for a source, $I_{q}<0$ for a sink, and $\sum_{sources}I_{q}=-\sum
_{sinks}I_{q}$). The conservation laws \ref{kirchhoff} alone do not uniquely
determine the flow in each pipe of the network. In many situations $i_{k}$
also derives from a potential function (electrical potential, pressure,
concentration, temperature,...) so that the potential difference $v_{k}$, the
flow rate $i_{k}$, and the resistance $r_{k}$ of pipe $k$ are related by Ohm's
law $v_{k}=r_{k}i_{k}$. In this case, the flow distribution is unique, and
each flow rate $i_{k}$ is an implicit function of the pipe lengths and pipe cross-sections.

Both the \textit{network geometry} (characterized by the pipe cross-sections
and pipe lengths) and \textit{topology} (the number of pipes and junctions,
and their specific arrangement) can be optimized in order to minimize $U$.
However, minimization must be done with some constraint on the pipe
cross-sections (otherwise, the optimization problem would be trivial: any
network connecting the sources to the sinks with infinitely large pipes would
be a solution). Here, a global constraint $C_{n}=$ $\sum_{k}l_{k}s_{k}^{n}$ on
the total volume ($n=1$) or total surface area ($n=1/2$) of the network is
considered. Such a global constraint is less restrictive than the local
constraint used in other recent studies on optimal networks
\cite{Banavar,Dougherty}, where every pipe cross-section is fixed.

Let us now prove that, under the assumptions above, the network that minimizes
$U$ is loopless. Let us start with a network of given topology, whose
geometrical parameters (pipe cross-sections and pipe lengths) are adjusted to
minimize the dissipation rate (while preserving $C_{n}$). Indeed, optimization
of the network geometry has been studied in a previous work \cite{Durand}:
first, when pipe cross-sections are adjusted, the flow rate $i_{k}$ carried by
each pipe in the network scales with its cross-sectional area $s_{k}$ as:%
\begin{equation}
\left\vert i_{k}\right\vert =\kappa Is_{k}^{\left(  m+n\right)  /2}%
,\label{currents}%
\end{equation}
where $I$ is the total flow rate through the network ($I=\sum_{sources}%
I_{q}=-\sum_{sinks}I_{q}$), and $\kappa$ is a parameter that depends on $m$,
$n$, and the geometry and topology of the network. Thus, the dissipation rate
in this network is:%
\begin{equation}
U=\left(  \kappa I\right)  ^{2}C_{n}.\label{energy}%
\end{equation}
Then, pipe lengths can also be adjusted in order to minimize $U$, while
preserving $C_{n}$ (according to Eq. \ref{energy}, this is equivalent to
minimizing $\kappa$, which still depends on the pipe lengths). Actually,
coordinates of the additional nodes are the appropriate independent
optimization parameters. When these coordinates can be freely adjusted, it has
been shown \cite{Durand} that the following vector balance is also satisfied
at every additional node of the network with optimized cross-sections
\textit{and} node locations:%
\begin{equation}%
{\displaystyle\sum\limits_{k}}
s_{k}^{n}\mathbf{e}_{k}=\mathbf{0,}\label{angles}%
\end{equation}
where $\mathbf{e}_{k}$\ is the outward-pointing unit vector along each
adjoining pipe. No such geometrical rule can be established when the locations
of the additional nodes must be chosen from a set of nodes (such as points of
a grid, or some particular cities of a country). It must also be noted that
Eqs. \ref{currents} and \ref{angles} are \textit{necessary} conditions for the
minimum of $U$ with respect to the geometrical parameters.

\begin{figure}[h]
\includegraphics[width=8cm]{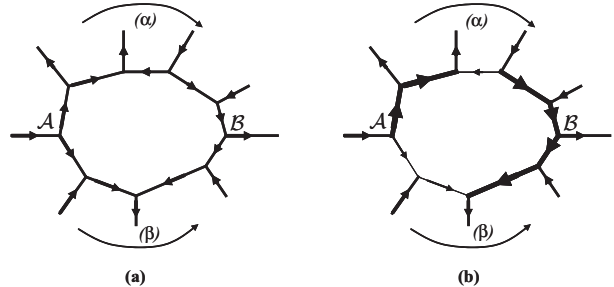}\caption{Shift of material in a loop
of the network. (a): the original loop, where flow directions in each pipe are
indicated with arrows. (b): the same loop, where the cross-sectional area of a
pipe is increased when the direction of its carried flow is $\mathcal{A}%
\rightarrow\mathcal{B}$ along path $\left(  \alpha\right)  $ or $\mathcal{B}%
\rightarrow\mathcal{A}$ along path $\left(  \beta\right)  $, and decreased
otherwise. The other cross-sectional areas in the network remain unaltered.}%
\label{figure1}%
\end{figure}

Suppose that this network, which satisfies Eq. \ref{currents} (and possibly
Eq. \ref{angles}), contains loops. Let us show that, from this original
network, a new loopless network with a lower dissipation rate (and with a same
value of $C_{n}$) can be built. Consider an arbitrary loop in this network. To
go from a given junction $\mathcal{A}$ to another junction $\mathcal{B}$ of
this loop, there are two different paths, noted $\left(  \alpha\right)  $ and
$\left(  \beta\right)  ,$ as depicted on Fig. \ref{figure1}a. Let us make a
shift of material, in such a way that flows in path $\left(  \alpha\right)  $
tend to be strengthened in one direction (say $\mathcal{A}$ to $\mathcal{B}$)
and flows in path $\left(  \beta\right)  $ tend to be strengthened in the
opposite direction ($\mathcal{B}$ to $\mathcal{A}$). That is, the new
cross-sectional areas $s_{k}^{\prime}$ in the loop are defined as:
$s_{k}^{\prime\left(  m+n\right)  /2}=s_{k}^{\left(  m+n\right)  /2}\pm
s_{0}^{\left(  m+n\right)  /2}$ with, for path $\left(  \alpha\right)  $, a
plus sign if flow rate in pipe $\left(  i,j\right)  $ is in direction
$\mathcal{A}\rightarrow\mathcal{B}$ and a minus sign if the flow rate is in
opposite direction, while signs are inverted for path $\left(  \beta\right)  $
(see Fig. \ref{figure1}b). $s_{0}$ is a positive number smaller than any
cross-sectional area $s_{k}$ of the original loop. Cross-sections outside the
loop remain unaltered ($s_{k}^{\prime}=s_{k}$). Note that flows in a loop
cannot turn all clockwise, or counterclockwise (otherwise, the potential
difference $V_{\mathcal{A}}-V_{\mathcal{B}}$ between nodes $\mathcal{A}$ and
$\mathcal{B}$, and thus the flow rates in the loop, would trivially be zero).
This guarantees that the cross-sectional areas of the loop did not all
simultaneously increase (or decrease).

Such a variation of cross-sectional areas implies a redistribution of flows in
the entire network. Let $\left\{  i_{k}^{\prime}\right\}  $ be the new
distribution of flow rates satisfying Eq. \ref{kirchhoff} and Ohm's law,
$r_{k}^{\prime}=\rho l_{k}/s_{k}^{\prime m}$ the new resistances, and
$U^{\prime}=\sum_{k}r_{k}^{\prime}i_{k}^{\prime2}$ the new dissipation rate.
Although we do not know the values of the new flow rates, an upper bound on
the new dissipation rate $U^{\prime}$ can be established, using Thomson's
principle \cite{Doyle,Jeans}. Consider a network of given resistances
$r_{k}^{\prime}$ that connects the sources to the sinks. Thomson's principle
states that -- among all possible flow distributions $\left\{  j_{k}\right\}
$ which satisfy the equations of conservation \ref{kirchhoff} -- the actual
flow distribution (i.e.: the one deriving from a potential function and
satisfying Ohm's law) is the one that makes the function $\sum_{k}%
r_{k}^{\prime}j_{k}^{2}$ an absolute minimum. Let us consider in particular
the flow distribution defined as: $j_{k}=i_{k}+i_{0}$ along path $\left(
\alpha\right)  $, $j_{k}=i_{k}-i_{0}$ along path $\left(  \beta\right)  $, and
$j_{k}=i_{k}$ for any pipe outside the loop. $i_{0}$ is some positive number,
and $\left\{  i_{k}\right\}  $ is the actual distribution in the original
network, the sign of $i_{k}$ being (re)defined in both paths as positive if
directed from $A$ to $B$. \ The distribution $\left\{  j_{k}\right\}  $
satisfies conservation equations \ref{kirchhoff}, since the distribution
$\left\{  i_{k}\right\}  $ does. Besides, by choosing $i_{0}=\kappa
Is_{0}^{\left(  m+n\right)  /2}$ and using Eq. \ref{currents}, the flow rate
distribution $j_{k}$ can be rewritten: $j_{k}=sgn\left(  i_{k}\right)  \kappa
Is_{k}^{\prime\left(  m+n\right)  /2}$. Thus, according to Thomson's
principle:
\begin{equation}
U^{\prime}\leq\left(  \kappa I\right)  ^{2}C_{n}^{\prime}, \label{inequality2}%
\end{equation}
with $C_{n}^{\prime}=$ $\sum_{k}l_{k}s_{k}^{\prime n}$. Let us now compare the
new value of pipe volume/surface area $C_{n}^{\prime}$ with the original value
$C_{n}$. This can be done by studying the variation of $C_{n}^{\prime}$ with
$s_{0}$. The derivative of this function with respect to $x=s_{0}^{\left(
m+n\right)  /2}$ is:
\begin{align}
\frac{\partial C_{n}^{\prime}}{\partial x}  &  =\frac{2n}{m+n}\sum_{path\text{
}\alpha}l_{k}\left(  s_{k}^{\left(  m+n\right)  /2}+x\right)  ^{\left(
n-m\right)  /\left(  n+m\right)  }\\
&  -\frac{2n}{m+n}\sum_{path\text{ }\beta}l_{k}\left(  s_{k}^{\left(
m+n\right)  /2}-x\right)  ^{\left(  n-m\right)  /\left(  n+m\right)
}.\nonumber
\end{align}
Since $m\geq n$, $\frac{\partial C_{n}^{\prime}}{\partial x}$ is a decreasing
function of $x$, and then:
\begin{equation}
\frac{\partial C_{n}^{\prime}}{\partial x}\leq\left(  \frac{\partial
C_{n}^{\prime}}{\partial x}\right)  _{x=0}. \label{inequality3}%
\end{equation}
Therefore, if the bound in inequality \ref{inequality3} is negative,
$C_{n}^{\prime}$ is a decreasing function of $x$. Using Eqs \ref{resistance}
and \ref{currents}, \ this bound can be rewritten as:
\begin{equation}
\left(  \frac{\partial C_{n}^{\prime}}{\partial x}\right)  _{x=0}=\frac
{2n}{m+n}\left(  \sum_{path\text{ }\alpha}r_{k}\left\vert i_{k}\right\vert
-\sum_{path\text{ }\beta}r_{k}\left\vert i_{k}\right\vert \right)  .
\label{bound}%
\end{equation}

We could have chosen to reinforce flows in direction $\mathcal{B}%
\rightarrow\mathcal{A}$ in path $\left(  \alpha\right)  $, and $\mathcal{A}%
\rightarrow\mathcal{B}$ in path $\left(  \beta\right)  $ instead, which comes
to swapping $\left(  \alpha\right)  $ and $\left(  \beta\right)  $ in the
calculations above. Inequalities \ref{inequality2} and \ref{inequality3} would
still be satisfied for this new shift of material, but this time with an
opposite sign for $\left(  \frac{\partial C_{n}^{\prime}}{\partial x}\right)
_{x=0}$ (see Eq. \ref{bound}). So, necessarily $\left(  \frac{\partial
C_{n}^{\prime}}{\partial x}\right)  _{x=0}\leq0$ for one of the two shifts,
and $C_{n}^{\prime}$ is a decreasing function of $s_{0}$ for this particular
shift, implying that $C_{n}^{\prime}\left(  s_{0}\right)  \leq C_{n}^{\prime
}\left(  0\right)  =C_{n}$. From Eqs. \ref{energy} and \ref{inequality2}, we
obtain that the corresponding dissipation rate $U^{\prime}$ is also lower:
$U^{\prime}\left(  s_{0}\right)  \leq U$ \cite{note1b}.

In a further step, the total volume/surface area can be increased up to its
original value $C_{n}$ by increasing any cross-sectional areas in the network.
This will imply a further decrease in $U$ \cite{note2}. Thus, we find a small
perturbation of the cross-sections such that the dissipation is reduced for a
fixed value of $C_{n}$ \cite{note4}. The reasoning above can be applied with
increasingly large values of $s_{0}$, until eventually one of the pipes in the
loop has a zero cross-sectional area, and so one of the paths is cut off.
Possible dead branches can be removed, the equivalent material being shifted
to the rest of the network by increasing any other cross-sectional areas
again, so that the constraint stays at its initial value while the dissipation
rate is subjected to a further decrease. Finally, the whole procedure can be
repeated to eliminate all the duplicate paths until there are no loops in the
network. The argument holds even in case of overlapping loops (that is, loops
having pipes in common), and more generally for any topology of the original
network. Therefore, it comes that the architecture of the network that
minimizes $U$ is loopless. Note that condition \ref{angles} is not used
throughout the reasoning, so the demonstration is valid whether or not the
positions of additional nodes can be freely adjusted.

It must be mentioned that the absence of loops in the least dissipative
network has already been conjectured (without proof) in the particular case of
a constrained total volume \cite{Bejan}. A similar result has also been
obtained in other studies on optimal networks: Banavar \textit{et al.}
\cite{Banavar} analyzed the flow rate distribution minimizing the cost
function $\sum_{k}w_{k}\left\vert i_{k}\right\vert ^{\gamma}$. They showed
that the flow pattern formed by this distribution contains no loops if
$0\leq\gamma<1$, and contains loops if $\gamma>1$. In that study however, both
the network topology and the weight $w_{k}$ of every link are set. The
optimization parameters are the flow rates $i_{k}$, subject to the
conservation laws Eq. \ref{kirchhoff} only (they do not necessarily derive
from a potential function and obey Ohm's law). This optimization problem is
then very different than the one analyzed in the present letter. In a
different study, Xue \textit{et al.} \cite{Dougherty} showed that the network
minimizing the cost function $\sum_{k}l_{k}\left\vert i_{k}\right\vert
^{\gamma}$ is also loopless when $0\leq\gamma<1$ \cite{note3}. In that study,
both the network topology and the pipe lengths are free to be adjusted, but
the cross-sections are set to $1$; instead of a global constraint on the total
pipe volume/surface area, Xue \textit{et al.} consider a more restrictive
constraint on every pipe cross-section. Therefore, the absence of loops in the
network minimizing $U$ and preserving $C_{n}$ cannot be deduced from that
study. Indeed, using Eqs. \ref{currents} and \ref{energy}, the dissipation
rate of a network with optimized cross-sections can be rewritten in terms of
the pipe lengths and the flow rates alone: $U\ =\left(  \kappa I\right)
^{\delta}\sum_{k}l_{k}\left\vert i_{k}\right\vert ^{\gamma}$, where
$\delta=2m/\left(  m+n\right)  $ and $\gamma=2n/\left(  m+n\right)  \leq1$.
The prefactor $\kappa$ in this expression is a parameter dependent on the pipe
lengths. Thus, $U$ differs clearly from the cost function studied in
\cite{Dougherty}.

\begin{figure}[h]
\includegraphics[width=8cm]{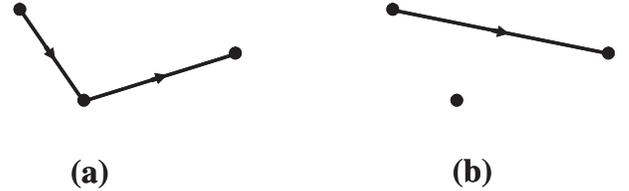}\caption{(a): the two adjoining pipes
of a two-fold junction carry flows in opposite directions in order to satisfy
flow rate conservation. (b): the two adjoining pipes can be favorably replaced
with a straight one: since the total pipe length is shortened, the dissipation
rate will be decreased for a fixed value of $C_{n}$.}%
\label{figure2}%
\end{figure}

Let us now show that the number of additional nodes is at most $N-2$ in the
optimal network, where $N$ $\ $is the total number of sources plus sinks. This
result limits the number of possible topologies for the optimal network.
Suppose first that the optimal network is a \textit{connected} loopless
network (or tree). According to Euler's formula \cite{Rivier}, a tree has one
more node than it has links. So, the number of links in a network with $A$
additional nodes is $N+A-1$. Since each link has two ends, the number of
\textquotedblleft incident lines\textquotedblright, summed over all the nodes,
is $2\left(  N+A-1\right)  $. This number can be evaluated differently: let
$N_{p}$ be the number of sources or sinks with $p$ incident lines. Since each
source or sink is linked to the rest of the tree, the smallest value of $p$
for which $N_{p}$ has a nonzero value is $1$, so $\sum_{p\geq1}N_{p}=N$.
Similarly, let $A_{p}$ be the number of additional nodes with $p$ incident
lines. By definition, two-fold junctions can exist only if its two links are
not parallel. Therefore, such junctions cannot exist in a network satisfying
Eq. \ref{angles}. Two-fold junctions could \textit{a priori} exist if their
positions are chosen from a set of nodes. However, they can be favorably
(i.e.: with no increase of $U$ and $C_{n}$) removed and their two adjoining
pipes replaced with a straight one, as depicted in Fig. \ref{figure2}. Thus,
the smallest value of $p$ for which $A_{p}$ is not zero is $p=3$ in both
cases, and the total number of incident lines is: $\sum_{p\geq1}pN_{p}%
+\sum_{p\geq3}pA_{p}$. Comparing these two expressions for the number of
incident lines, and considering that $\sum_{p\geq1}pN_{p}\geq N$ and
$\sum_{p\geq3}pA_{p}\geq3A$, it appears that:
\begin{equation}
A\leq N-2, \label{Ineq}%
\end{equation}
as was to be proven. When both the number of sources and the number of sinks
are strictly larger than $1$, the optimal network might be disconnected.
However, using the reasoning above on each of the trees that constitute the
optimal network, it comes that the inequality \ref{Ineq} is still satisfied.

Because of the broad definition of the flow resistance (Eq. \ref{resistance}),
the results presented in this letter can be applied in various situations. For
instance, the $m=1$ case corresponds to electrical current in wires, liquid
flow in porous conducts, mass or heat diffusion in bars (provided that for the
latter, the bar lateral surface is insulated). The $m=2$ case corresponds to
the laminar Poiseuille flow in hollow pipes. Minimization can be done for a
fixed lateral surface area ($n=1/2$) if one wants to save the material
required to build the hollow pipes, or for a fixed volume ($n=1$), if one
wants to preserve the amount of liquid flowing through the network.
Furthermore, these results may also explain the tree-like structure of the
circulatory system of various living organisms \cite{McMahon,Banavar}.

Unfortunately, the results presented in this letter do not give insights into
the method of building the optimal network practically, or even into the
uniqueness of such an optimal network. In fact, as for the Steiner tree
problem - which consists in finding the tree of minimal length interconnecting
a set of given points - this problem is likely to be NP-hard, meaning that the
solution cannot be found without an exhaustive search of all the possible
topologies. However, the NP-hardness does not exclude the possibility of
establishing basic properties on the geometry and topology of Steiner trees
\cite{Gilbert}. Similarly, we were able to address features on the structure
of pipe networks minimizing the total dissipation rate under a global
constraint. Specifically, the upper bound on the number of additional nodes
restricts the number of possible topologies for the optimal network(s). These
results make possible the conception of efficient algorithms for computing the
optimal pipe network problem \cite{Dougherty}. In many situations however, the
capacity of the network to resist random injuries may also play a key role in
its design. Obviously, a reticulate network containing redundant paths is more
adapted than an arborescent one for that purpose. Therefore, it is sometimes
essential to look for a compromise between optimization of flow and robustness
of the network.

The author thanks B. Abou, S. Durand, T.\ Forth and A. Rabodzey for careful
reading of the manuscript.


\begin{thebibliography}{99}
\bibitem {note1b}Note that if all flows in both paths of the loop point
towards same direction -- say $\mathcal{A}$ to $\mathcal{B}$ -- in the
original network, then $\left\vert i_{k}\right\vert =i_{k}$. The sum of
$r_{k}i_{k}$ along each path is then equal to $V_{\mathcal{A}}-V_{\mathcal{B}%
}$, so that $\left(  \frac{\partial C_{n}^{\prime}}{\partial x}\right)
_{x=0}=0$. Thus, either one of the shifts from $\left(  \alpha\right)  $ to
$\left(  \beta\right)  $ or from $\left(  \beta\right)  $ to $\left(
\alpha\right)  $ leads to a decrease in the dissipation rate in that specific case.
\bibitem {note3}This result holds for any flow distribution satisfying the
conservation laws Eq. \ref{kirchhoff} (the flow rates $i_{k}$ are not
optimization parameters, and they do not necessarily obey Ohm's law).
\end{thebibliography}
\end{document}